\documentclass[letterpaper,twoside,onecolumn,final]{article}

\usepackage[UKenglish]{babel}
\usepackage{geometry}
\usepackage{setspace}

\usepackage[]{cite}

\usepackage[fontsize=12.4]{fontsize}
\usepackage{graphicx}
\usepackage{float}

\usepackage{amsmath,amssymb,amsfonts,
            amsthm,mathtools,mathrsfs,
            stmaryrd}
\usepackage{braket}                 														%
\usepackage[retainorgcmds]{IEEEtrantools} 													%
\usepackage{mleftright}             														%
\mleftright                         %

\usepackage{booktabs,tabularx,makecell}        												%
\usepackage[labelformat=simple]{subcaption}                     							%
\usepackage{siunitx,pgf}     														%

\usepackage{overpic}                                                                        %
\usepackage[norefs,nocites,ignoreunlbld]{refcheck}
\usepackage{hyperref}               														%
\hypersetup{
	unicode,
	hidelinks,
	urlcolor=cyan,
	bookmarksopen=true,
	bookmarksopenlevel=1,
	pdfstartview={XYZ null null 1},
	pdfpagemode={UseOutlines},
	pdfpagelayout={OneColumn},
	pdfpagemode=FullScreen,
  pdfinfo={ Creator={}, Producer={}, ModDate={...}, CreationDate={...} }
}
\usepackage{zref-clever,csquotes,xpatch}
\AddToHook{env/IEEEeqnarray/begin}{%
  \zcsetup{ currentcounter = equation }%
}
\usepackage{zref-titleref}

\AddToHook{begindocument}{\zcsetup{countertype={subfigure=summfigure}}}

\zcsetup{cap,noabbrev,S,
}

\usepackage{zref-xr}
\zexternaldocument{supp_report}
\zxrsetup{tozreflabel=false,toltxlabel=true}
\zexternaldocument*{supp_report}
\zcLanguageSetup{UKenglish}{
  type = suppsection ,
    Name-sg = SI Section ,
    name-sg = si,
    Name-pl = SI Sections ,
    name-pl = si,
  type = suppfigure ,
    Name-sg = Figure SI,
    name-sg = si,
    Name-pl = Figures SI,
    name-pl = si,
  type = summfigure ,
    Name-sg = Figure,
    name-sg = si,
    Name-pl = Figures,
    name-pl = si,
  type = supptable ,
    Name-sg = Table SI,
    name-sg = si,
    Name-pl = Tables SI,
    name-pl = si,
}

\usepackage{authblk}
\author[1,2]{Luis Vasquez*}
\author[1]{Kewei Sun}
\author[3]{Haibo Ma}
\author[1]{Maxim F. Gelin*}
\affil[1]{School of Science, Hangzhou Dianzi University, Hangzhou 310018, China}
\affil[2]{Graduate School of Computer and Information Sciences, Hosei University, Tokyo 184-8584, Japan}
\affil[3]{Key laboratory of colloid and interface chemistry, School of
          Chemistry and Chemical Engineering, Shandong University, Qingdao 266237, China}

\title{An \emph{Ab initio} Framework for Simulating Ultrafast Nonlinear Cavity Quantum Electrodynamics Spectra}

\date{*Email: maxim@hdu.edu.cn, 90027@hdu.edu.cn}

\makeatletter
\def\IEEElabelanchoreqn#1{\bgroup
	\def\@currentlabel{\p@equation\theequation}\relax
	\def\@currentHref{\@IEEEtheHrefequation}\label{#1}\relax
	\Hy@raisedlink{\hyper@anchorstart{\@currentHref}}\relax
	\Hy@raisedlink{\hyper@anchorend}\egroup}

\renewcommand{\@seccntformat}[1]{}
\makeatother
\renewcommand{\thesection}{\Alph{section}}
\makeatletter
\AddToHook{cmd/appendix/before}{%
    \def\cref@section@alias{appendix}
}

\renewcommand\appendix{\par
  \counterwithin*{equation}{section}
  \renewcommand{\thesection}{\Alph{section}}
  \renewcommand{\theequation}{\Alph{section}.\arabic{equation}}
  \setcounter{section}{0}%
  \setcounter{equation}{0}%
  \gdef\@chapapp{\appendixname}%
  \gdef\thechapter{\appendixname\@Alph\c@chapter}}
\makeatother

\begin{document}

\maketitle

\begin{abstract}
    In this letter we introduce a theoretical framework for the simulation of
    ultrafast transient absorption pump-probe spectra of molecular
    polaritons. We derive and implement the cavity quantum electrodynamics (QED)
    evolution equations of polaritonic states within the framework of the quasi-classical
    doorway-Window approximation, hereto referred cQUEDA \emph{/sikeda/}. %
    This framework uses outputs from mixed quantum-classical dynamics simulations in the absence of the cavity. Then
    by including cavity parameters accounting for cavity rate loss, coupling
    strength, and frequency detuning, we simulate transient absorption pump-probe spectra.
    Consequently, we address one of the main standing problems of cavity QED,
    the simulation of polaritons ultrafast dynamics and nonlinear optical properties.
    We demonstrate the performance of our method by computing the  ground-state bleach (GSB), stimulated emission (SE),
    and excited-state absorption (ESA) contributions of transient
    absorption pump-probe spectra of pyrazine strongly coupled to a cavity. 
    The cQUEDA is an on-the-fly computationally efficient framework  
    with low computer requirements: the pyrazine calculations, for example, took minutes on modern laptops.
    cQUEDA offers wide-ranging applicability and can be generalized
    to model diverse nonlinear spectroscopic signals and quantum optics responses.

\end{abstract}

\clearpage

\section{Introduction\zlabel{se:intro}}

One of the main areas of study in cavity quantum electrodynamics (QED) concerns
the emergent phenomena associated with coupled light-matter hybrid states,
known as polaritons. Polaritonic states underpin state-of-the-art applications,
like molecular sensing,\cite{Hutchison2012} remote control of chemical reactions,\cite{Biswas2025}
photoswitching,\cite{Kuttruff2023,Gernet2024} photonic quantum information,\cite{Cuevas2018}
and quantum computing.\cite{Barrat2024} Consequently, most experimental investigations
focus on the finely tuned control of polaritonic optical properties.

Recent advances in experimental techniques and device development permit us
to elucidate intricate polariton phenomena which are fundamental for modulating polariton-based photophysics
and photochemistry.\cite{Wang2021a,Zeng2023,Kuttruff2023,Bhuyan2023,McKillop2026}
With the development of those experimental techniques, also guidelines are
emerging to properly interpret the polaritonic spectra,\cite{Renken2021,Schwennicke2025}
as optical artifacts could arise without proper intervention.\cite{Piejko2024}
Those novel experimental techniques often rely on nonlinear femtosecond
spectroscopy for characterizing the time-resolved ultrafast dynamics and nonlinear
optical properties of polaritonic modes; the lower polariton (LP), the upper polariton
(UP), and the less understood optically inactive dark states.\cite{Khazanov2023,Pandya2022}

Among the emerging experimental techniques for characterizing polaritonic
systems, ultrafast transient absorption (TA) pump-probe (PP)
spectroscopy has been especially helpful for capturing spectral signatures
of cavity-induced dynamics.\cite{Virgili2011,Wu2022,McKillop2026,Rashidi2025}
This have had direct implications for solving the ongoing controversy on the
role of excited-state lifetimes in the tuning and modulation of photochemical
QED processes.\cite{Thomas2023,Piejko2024,Thomas2024,Schwartz2025,Thomas2025}
Recently, a state-of-the-art ultrafast TA PP investigation provided conclusive
evidence that dynamics are unlikely perturbed once molecules leave
the Franck-Condon region or relax into the incoherent reservoir, indicating
that QED responses on long timescales (\(>\)\qty{1}{\ps}) have to be
carefully analized.\cite{McKillop2026}

On the theoretical side, the polariton dynamics were comprehensively studied using several
approaches, \emph{e.g.}\ exact quantum dynamics,\cite{Hou_2024,Sun2022,Pino2018}
mixed quantum-classical (MQC),\cite{Zhou2022} master-equation methods,\cite{Mondal2025a}
and similar approaches (see reviews in Refs. \cite{Ruggenthaler2023,Mandal2023})
Analytical derivations
exist for most relevant polariton linear-response observables.\cite{YuenZhou2024}
Recently, a master-equation-based framework was developed for simulating
two-dimensional (2D) nonlinear spectra in the collective regime.\cite{Mondal2025}
Another notable study was focused on the simulation of transient photon absorption.\cite{Silva2020}

The appeal to simulate TA PP spectra is threefold: first, it can help to eliminate 
experimental artifacts by employing robust cavity QED principles;\cite{Renken2021,Piejko2024,Thomas2024,Nelson2024,McKillop2026}
second, cavity discretization elucidates phenomena emerging while changing
from the single-molecule to the collective regime;\cite{Silva2020,Chng2024,Dutta2024,Borges2025}
third, it provides access to spectroscopic signals beyond the reach of
state-of-the-art experiments and to features inaccessible to computational
linear spectroscopy, \emph{e.g.}\ intrinsic dynamics of the molecule and
cavity response, carrier and reservoir dynamics, and many-body nonequilibrium
effects.\cite{Sun2022,YuenZhou2024,Weight2025a}
However, despite recent developments, the simulation of nonlinear spectroscopic
responses of QED molecular systems remains challenging.

Leveraging MQC approaches, the doorway-window (DW) approximation for TA PP
spectra -- developed within the framework of the third-order response-function
formalism \cite{Mukamel95,DW1,DW2,DW3} -- has been used to simulate nonlinear
spectroscopic signals by \emph{ab initio} methods, \emph{e.g.}\ TA PP spectra,\cite{Xiang21} strong field PP
spectra,\cite{Guan2025,Zhang2026} pump-pump-probe signals, \cite{Bai2026} and
several others.\cite{Xiang2D,Xu2024,Sun2024,Pios2024} It is essential that the DW approximation enables the on-the-fly
simulation  of the three contributions to spectroscopic signals, namely
ground-state bleach (GSB), stimulated emission (SE), and excited-state
absorption (ESA).
Inspired by the development of the TA PP and strong field PP spectra within
the  DW approximation,\cite{Xiang21,Guan2025,Zhang2026} we extend the DW
framework to polaritonic systems starting from the effective
Jaynes-Cummings Hamiltonian.\cite{Jaynes1963} This produces the polaritonic
DW approximation, hereto referred to as cQUEDA \emph{/sikeda/}.
We demonstrate the functionality of our methodology by computing the cavity-QED 
GSB, SE, and ESA contributions to the TA PP spectra of a prototypical
polyatomic chromophore, pyrazine.

The structure of this letter is as follows. In \zcref{sse:4wm} we introduce the
physical basis of the methodology, while in \zcref{se:results} we discuss
the implementation of the polaritonic-DW methodology and the simulated
TA PP signals. Finally, in \zcref{se:conclutions} we discuss the applications,
perspectives and further developments of the methodology.

\section{Nonlinear Cavity QED spectra\zlabel{sse:4m}}

The quasi-classical DW approximation hinges on partitioning of the molecular
electronic states into three manifolds, \{0\}, \{I\},
and \{II\} where manifold \{0\} contains the electronic ground
state, manifold \{I\} contains the bright electronic states that the
pump pulse can access from the ground state, as well as
other electronic states intramolecularly coupled to them; and manifold \{II\}
comprises the electronic states that laser pulses can probe from manifold \{I\}.\cite{Gelin22}

To treat a polaritonic system, we add the cavity photonic mode to the
molecular states of manifold \{I\}. Hence, our polaritonic  Hamiltonian
assumes the form
\begin{IEEEeqnarray}{rCl}
    H & = &
    \begin{pmatrix}
        H_{0} & 0                   & 0                   & 0               \\
        0     & H_{\mathrm{I}}      & \Delta (t) & 0               \\
        0     & \Delta (t) & H_{\mathrm{c}}      & 0               \\
        0     & 0                   & 0                   & H_{\mathrm{II}}
    \end{pmatrix}.
    \IEEEeqnarraynumspace
    \zlabel{eq:big_cav}
\end{IEEEeqnarray}
Here $H_k$ is the molecular Hamiltonian in manifold $k$;
\(H_{\mathrm{c}} = \hslash\omega_{\mathrm{c}}\) is the
cavity Hamiltonian, \(\omega_{\mathrm{c}}\) is the cavity mode frequency,
and \(\Delta (\mathrm{t})\) is the time-dependent cavity-molecule
coupling. We consider coupling of the lowest excited states of the
molecule to the cavity and ignore the states in which both the molecule
and the cavity are excited (so called doubly-excited states).

Subsequently, we exploit the form of the Hamiltonian of \zcref{eq:big_cav}
and link this with the effective description of the molecular polariton system 
that accounts for polariton decay in a single excitation.\cite{Mandal2023}

\subsection{Cavity QED in the Doorway-Window approximation\zlabel{sse:4wm}}

In this section, we expand on the equations to calculate the nonlinear cavity QED
spectra within the DW approximation framework. For a detailed review of the
DW approximation, we refer the reader to Refs. \cite{Gelin2025,Mukamel95}.
\zcref{sec:qed_derive} details the derivation of cQUEDA.
Lastly, the terminology, the equations' syntax, and nomenclature
used in this section closely follow that of Ref. \cite{Taylor2025}

We begin by denoting the molecular adiabatic states in manifold \{I\} as
\(e\) and the corresponding electronic energies as \(V_{e}\).
The coupling to the cavity splits \(V_{e}\) into the doublet with the energies
\begin{IEEEeqnarray}{rCl}
    V_{e}^{(1)}(T) & = &  \frac{V_{e}(T)+\hslash \omega_{\textrm{c}}+ \Omega_{e}(T)}{2},
    \IEEEeqnarraynumspace
    \zlabel{eq:up_quant_E}
\end{IEEEeqnarray}
and
\begin{IEEEeqnarray}{rCl}
    V_{e}^{(2)}(T) & = &  \frac{V_{e}(T)+\hslash \omega_{\textrm{c}}- \Omega_{e}(T)}{2},
    \IEEEeqnarraynumspace
    \zlabel{eq:lo_quant_E}
\end{IEEEeqnarray}
where $T$ is the time specifying evolution of all quantities along the
quasi-classical trajectory. \(V_{e}^{(1)}(T)\) and
\(V_{\textrm{e}}^{(2)}(T)\) are separated by the so-called Rabi
frequency (\(\Omega_{e}\)),
\begin{IEEEeqnarray}{rCl}
    \Omega_{e}(T) & = & \sqrt{\Delta_{\omega}^{2} +
        4\gamma^{2}(T)\left|\mu_{ge}(T)\right|^{2}}.
    \IEEEeqnarraynumspace
    \zlabel{eq:rabi}
\end{IEEEeqnarray}
Here \(\Delta_{\omega} = \hslash \omega_{\textrm{c}} - V_{e}(T),\) represents
the cavity detuning. At resonance  \(\Delta_{\omega} = 0\).
\begin{IEEEeqnarray}{rCl}
    \gamma\left(T\right) &=& \gamma_{0}e^{-\kappa T},
    \IEEEeqnarraynumspace
    \zlabel{eq:coupling_t}
\end{IEEEeqnarray}
is a function that expresses the dipole coupling strength of the state
\(e\) to the cavity in terms of the transition dipole moment \(\mu_{ge}\). 
The time dependence of \(\gamma(T)\) is responsible for the finite
lifetime \(1/\kappa\) of the QED photonic-mode resonance, \(\kappa\) is
the cavity loss rate, and \(\kappa/\hslash \equiv \tau_{c}^{-1}\) where
\(\tau_{c}\) is the cavity lifetime.
For an ideal cavity, \emph{i.e.}\
\(\kappa \rightarrow 0\), \(\gamma\left(T\right) = \gamma_{0}\); therefore,
the photonic mode lives indefinitely on the simulation timescale.
In practical simulations, \(\gamma_{0}\) must be non-zero; otherwise,
\zcref{eq:coupling_t} leads to the singularity \(0/0\) which, however, has
a well defined limit for \(\gamma_{0} \rightarrow \infty\). In this latter
case, the light-matter coupling vanishes, returning us to the conventional
DW formulas.\cite{Gelin2025}

We also introduce the quantities
\begin{IEEEeqnarray}{rCl}
    \sigma_{e}\left(T\right)   &=& \Omega_{e}(T)+\hslash \omega_{\textrm{c}}-V_{e}(T).
    \IEEEeqnarraynumspace
    \zlabel{eq:cav_E}
\end{IEEEeqnarray}
and
\begin{IEEEeqnarray}{rCl}
    \cos^{2} \left( \phi\left(T\right) \right) = \frac{2\gamma^{2}(T) \left|\mu_{\textrm{ge}}(T)\right|^{2}}{\Omega_{e}(T)\sigma_{e}(T)},
    & \,\,\,& \sin^{2}\left( \phi\left(T\right) \right) = \frac{\sigma_{e}(T)}{2\Omega_{e}(T)}
    \nonumber
    \IEEEeqnarraynumspace
    \zlabel{eq:trigon_E}
\end{IEEEeqnarray}
where \(\sigma_{e}\left(T\right)\) is the light-matter cavity energy,
\(\cos^{2} \left( \phi\left(T\right) \right) \textrm{ and }
\sin^{2}\left( \phi\left(T\right) \right) \) are the parameters
specifying coupling to the cavity (see \zcref{sec:qed_derive}).

Finally, we define
\begin{IEEEeqnarray}{rCl}
    U_{eg}^{(1)}(T)=V_{e}^{(1)}(T)-V_{g}(T), & \,\,\,& U_{eg}^{(2)}(T)=V_{e}^{(2)}(T)-V_{g}(T).
    \nonumber
    \IEEEeqnarraynumspace
    \zlabel{eq:quantum_E}
\end{IEEEeqnarray}
These are the transition frequencies required for the calculation of the
TA PP spectra.

Now, we evaluate the nonlinear spectroscopic signal using the
standard formula \cite{Gelin2025}
\begin{IEEEeqnarray}{rCl}
    I_{PP}^{int}(T,\omega_{pr})  &\sim&  \langle D_{0} \left(\omega_{pu},\boldsymbol{R}_{g},\boldsymbol{P}_{g}\right)
    W_{0}^{int}\left(\omega_{pr},\boldsymbol{R}_{g}(T),\boldsymbol{P}_{g}(T)\right)\rangle \nonumber \\
    & & + \langle D_{\mathrm{I}}(\omega_{pu},\boldsymbol{R}_{g},\boldsymbol{P}_{g})W_{\mathrm{I}}^{int}(\omega_{pr},\boldsymbol{R}_{e}(T),\boldsymbol{P}_{e}(T))\rangle \nonumber \\
    & & - \langle D_{\mathrm{I}}(\omega_{pu},\boldsymbol{R}_{g},\boldsymbol{P}_{g})W_{\mathrm{II}}^{int}(\omega_{pr},\boldsymbol{R}_{e}(T),\boldsymbol{P}_{e}(T))\rangle.
    \IEEEeqnarraynumspace
    \zlabel{eq:S1a-1}
\end{IEEEeqnarray}
Here the doorway functions are defined as 
\begin{IEEEeqnarray}{rCl}
    D_{0}(\omega_{pu},\boldsymbol{R}_{g},\boldsymbol{P}_{g}) & = &
    \sum_{e}A_{eg}(\omega_{pu},0,\boldsymbol{R}_{g}) \left|\mu_{ge}(\boldsymbol{R}_{g})\right|^{2} \rho_{g}^{Wig}(\boldsymbol{R}_{g},\boldsymbol{P}_{g}),
    \IEEEeqnarraynumspace
    \zlabel{eq:dg1-1}
\end{IEEEeqnarray}
and
\begin{IEEEeqnarray}{rCl}
    D_{\mathrm{I}}(\omega_{pu},\boldsymbol{R}_{g},\boldsymbol{P}_{g}) & = &
    A_{eg}(\omega_{pu},0,\boldsymbol{R}_{g}) \left|\mu_{ge}(\boldsymbol{R}_{g})\right|^{2} \rho_{g}^{Wig}(\boldsymbol{R}_{g},\boldsymbol{P}_{g}),
    \IEEEeqnarraynumspace
    \zlabel{eq:de1-1}
\end{IEEEeqnarray}
where
\begin{IEEEeqnarray}{rCl}
    A_{eg}(\omega_{pu},0,\boldsymbol{R}_{g}) & = &
    E_{pu}^{2}(\omega_{pu}-U_{eg}^{(1)}(0,\boldsymbol{R}_{g}))\cos^{2}(\phi(0,\boldsymbol{R}_{g})) \nonumber \\
    & & + E_{pu}^{2}(\omega_{pu}-U_{eg}^{(2)}(0,\boldsymbol{R}_{g}))\sin^{2}(\phi(0,\boldsymbol{R}_{g})).
    \IEEEeqnarraynumspace
    \zlabel{eq:spect_energy_d}
\end{IEEEeqnarray}
The window functions are defined as follows:
\begin{IEEEeqnarray}{rCl}
    W_{0}^{int}(\omega_{pr},\boldsymbol{R}_{g}(T),\boldsymbol{P}_{g}(T)) & = &
    \sum_{e}B_{eg}(\omega_{pr},T,\boldsymbol{R}_{g}(T)) \left|\mu_{ge}(\boldsymbol{R}_{g}(T))\right|^{2},
    \IEEEeqnarraynumspace
    \zlabel{eq:wg1-1}
\end{IEEEeqnarray}
\begin{IEEEeqnarray}{rCl}
    W_{\mathrm{I}}^{int}(\omega_{pr},\boldsymbol{R}_{e}(T),\boldsymbol{P}_{e}(T)) & = &
    B_{e(T)g}(\omega_{pr},T,\boldsymbol{R}_{e}(T)) \left|\mu_{ge(T)}(\boldsymbol{R}_{e}(T))\right|^{2},
    \IEEEeqnarraynumspace
    \zlabel{eq:we1-1}
\end{IEEEeqnarray}
and
\begin{IEEEeqnarray}{rCl}
    W_{\mathrm{II}}^{int}(\omega_{pr},\boldsymbol{R}_{e}(T),\boldsymbol{P}_{e}(T)) & = &
    \sum_{f}B_{e(T)f}(\omega_{pr},T,\boldsymbol{R}_{e}(T)) \left|\mu_{e(T)f}(\boldsymbol{R}_{e}(T))\right|^{2}
    \IEEEeqnarraynumspace
    \zlabel{eq:wf1-1}
\end{IEEEeqnarray}
where 
\begin{IEEEeqnarray}{rCl}
    B_{eg}(\omega_{pr},T,\boldsymbol{R}_{g}(T)) & = &
    E_{pr}^{2}(\omega_{pr}-U_{eg}^{(1)}(T,\boldsymbol{R}_{g}(T)))\cos^{2}(\phi(T,\boldsymbol{R}_{g}(T))) \nonumber \\
    & & + E_{pr}^{2}(\omega_{pr}-U_{eg}^{(2)}(T,\boldsymbol{R}_{g}(T)))\sin^{2} (\phi(T,\boldsymbol{R}_{g}(T))),
    \IEEEeqnarraynumspace
    \zlabel{eq:spect_energy_w0}
\end{IEEEeqnarray}
\begin{IEEEeqnarray}{rCl}
    B_{e(T)g}(\omega_{pr},T,\boldsymbol{R}_{e}(T)) & = &
    E_{pr}^{2}(\omega_{pr}-U_{e(T)g}^{(1)}(T,\boldsymbol{R}_{e}(T)))\cos^{2}(\phi(T,\boldsymbol{R}_{e}(T))) \nonumber \\
    & & + E_{pr}^{2}(\omega_{pr}-U_{e(T)g}^{(2)}(T,\boldsymbol{R}_{e}(T)))\sin^{2}(\phi(T,\boldsymbol{R}_{e}(T))),
    \IEEEeqnarraynumspace
    \zlabel{eq:spect_energy_wI}
\end{IEEEeqnarray}
\begin{IEEEeqnarray}{rCl}
    B_{e(T)f}(\omega_{pr},T,\boldsymbol{R}_{e}(T)) & = &
    E_{pr}^{2}(\omega_{pr}-U_{ef}^{(1)}(T,\boldsymbol{R}_{e}(T)))\cos^{2}(\phi(T,\boldsymbol{R}_{e}(T)))  \nonumber \\
    & & + E_{pr}^{2}(\omega_{pr}-U_{ef}^{(2)}(T,\boldsymbol{R}_{e}(T)))\sin^{2}(\phi(T,\boldsymbol{R}_{e}(T))).
    \IEEEeqnarraynumspace
    \zlabel{eq:spect_energy_wII}
\end{IEEEeqnarray}

Having introduced the methodology for simulating cavity QED
TA PP signals within the semiclassical DW approximation, we discuss the
underlying approximations. 
Along with the standard DW approximations introduced in the trajectory DW
methodology (non-overlapping pump and probe pulses, quasi-classical
approximation)\cite{Gelin2025} we assume that the wavepacket dynamics
during \(T\) occurs without the influence of the cavity.
This assumption is justified if the cavity mode lifetime is short on
the molecular dynamics timescale. Hence our methodology can be applied to
short-lived cavity modes with the lifetimes up to dozen of \unit{\fs},
depending on the cavity properties, coupling strength and the number of
molecules in the cavity.\cite{Ojambati2019,Bhuyan2023}
We plan to relax this approximation in a future work by propagating
the wavepacket in the coupled light-matter states of manifold \(\{I\}\).

For polaritonic pyrazine simulations, we assume a brief coupling to the cavity.
Consequently, our methodology is justified and appropriate in
certain scenarios: (i) it applies to short-lived cavity resonances that
exist exclusively during the pump pulse. In this case, the cavity does
not affect the window function, which reduces to the standard form.
(ii) One begins from the matter state \(e\). Therefore, our approximation
remains valid for arbitrary cavity-mode lifetime at short \(T\), until
the wavepacket hops to the photonic state. (iii) If the wavepacket hops
to the photonic state, it does not
contribute to the signal because the corresponding transition dipole moments
is zero. Thus, the impact of the photonic state at intermediate \(T\)
may be regarded as a loss channel and described by an exponential decay
of the pump-probe intensity with \(T\) (correlation function in
\zcref{eq:coupling_t}).

Despite the limitations discussed above, the cQUEDA methodology
provides microscopic insight into cavity QED phenomena of current interest
such as polariton relaxation. The underlying DW expressions are analytical
and implicitly include cavity-mode lifetime. The latter is sometimes
neglected in polaritonic trajectory surface hopping
simulations.\cite{SangiogoGil2024} Our methodology,
however, captures the main physical processes in polaritonic systems with few
parameters, namely the polaritonic mode frequency \(\omega_{\mathrm{c}}\), decay rate \(\kappa\),
and the coupling strength \(\gamma_{0}\).
This leads us to two important points in the development of the methodology: (i)
the formulas for the DW functions are analytical. That preserves the spirit and
simplicity of the original trajectory DW methodology;\cite{Gelin2025}
(ii) owing to this simplicity, the methodology can be further refined 
and extended to suit other systems or situations.
One particularly interesting future development of the methodology is to
link trajectory simulations to experimental data in order to simulate
cavity QED spectra. Recent efforts have sought to connect model Hamiltonians
with experimental observables.\cite{Barnes2024} Knowing that the upper
polariton decays faster than the lower polariton, one could replace, for example,
the single exponential decay in \zcref{eq:coupling_t} with a biexponential
decay to achieve a better description of polariton relaxation.\cite{Barnes2024}.

Finally, the \(T\)-dependence in our simulations arises from quasi-classical
nonadiabatic simulations
(see simulation details in \zcref{se:simulations}). Consequently, the
cQUEDA accounts for nonadiabatic electron-nuclear couplings. 
Thus, by leveraging the DW approximation, the methodology developed here can,
in principle, track polariton relaxation and propagation.

\section{Cavity QED spectra\zlabel{se:results}}

\subsection{Computational Aspects\zlabel{sse:qed_implement}}

We implemented the cQUEDA in the package \texttt{WaveMixings.jl v0.8.0}
.\cite{Vasquez2026}
\texttt{WaveMixings.jl} is an open-source numerical implementation of the
DW methodology, written in the Julia programming language and 
maintained by the authors. For the cavity QED spectra calculations,
we distinguish between the cavity QED parameters (\(\omega_{\mathrm{c}}\),
\(\kappa\) and \(\gamma_{0}\)) and the DW parameters
(pulse duration \(\tau \), pump-pulse carrier  frequency \(\omega_{\textrm{pu}}\),
and probe-pulse carrier frequencies \(\omega_{\textrm{pr}}\)). For the purpose
of our calculations, we take
\(\tau = \qty{5.0}{\fs}\), \(\omega_{\textrm{pu}} = \qty{5.2}{\eV}\), and
\(\omega_{\textrm{pr}}\) in the range from \qtyrange{3.0}{7.0}{\eV} in
steps of \qty{0.02}{\eV}. On the other hand, we vary the cavity QED
parameters. Our final parameter is the pulse envelope, which we
assign a homogeneous Lorentzian shape based on analyses of polariton
absorption spectra.\cite{Houdre1996,Schwennicke2025}
For the cQUEDA spectra calculations and data analyses,
we used Julia \texttt{1.13.0-rc3} on a Intel i5-12500H platform equipped
with 16 GB of RAM.\cite{Bezanson2017,Bezanson2018}

Quasi-classical nonadiabatic simulations necessary for the spectra calculation
were performed in accordance with the simulation details in the
\zcref{se:simulations}. We confirmed through a convergence test that
the reported number of quasi-classical trajectories yields spectrally
converged results. Details of the spectral convergence tests are provided
in the \zcref{se:convergence}.

\subsection{Results\zlabel{sse:qed_results}}

\zcref{fig:cqed_signals} shows the computed cavity QED TA PP signals:
\zcref{fig:cav_gs} for GSB, \zcref{fig:cav_se} for SE, \zcref{fig:cav_es}
for ESA, and \zcref{fig:cav_tt} for the total signal.
All spectral plots confirm the existence of two polaritonic eigenstates representing
the LP and UP brances.
Interestingly, the LP exhibits significantly higher intensity than the UP,
despite the coupling occurring at the resonance frequency. We assign the
observed polariton Rabi splitting to the difference between UP and LP
intensity peaks in the SE contributing signal, \(\qty{0.949}{\eV}\).
Although the photoluminescence spectrum normally depicts Rabi splitting
experimentally, we refer here to the SE signal, as both processes involve
the same underlying excited-state radiative pathway and therefore yield
comparable results.

\begin{figure}[htbp!]
    \centering
    \begin{subcaptionblock}{.49\columnwidth}
        \centering
        \subcaptionlistentry{\(\omega_{\mathrm{c}} = 0\), \(\kappa = 10^{-6}\) and \(\gamma_{0} = 10^{-6} \)}
        \zlabel{fig:cav_gs}
		\begin{overpic}[width=\textwidth]{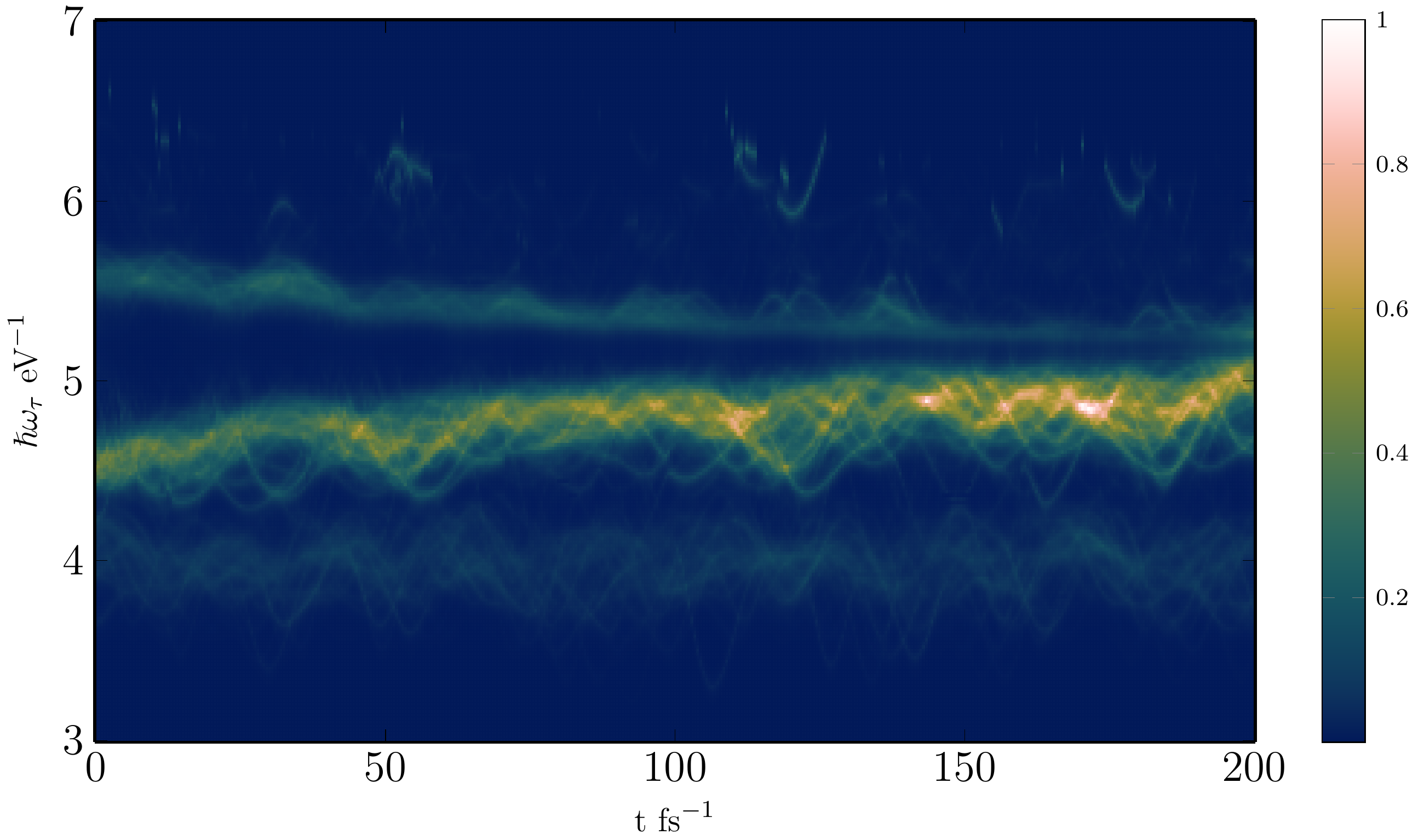}
			\put(10,52){\color{white}\captiontext*{}}
		\end{overpic}
    \end{subcaptionblock}
    \begin{subcaptionblock}{.49\columnwidth}
        \centering
        \subcaptionlistentry{\(\kappa = 10^{2}\) and \(\gamma_{0} = 0.011 \)}
        \zlabel{fig:cav_se}
		\begin{overpic}[width=\textwidth]{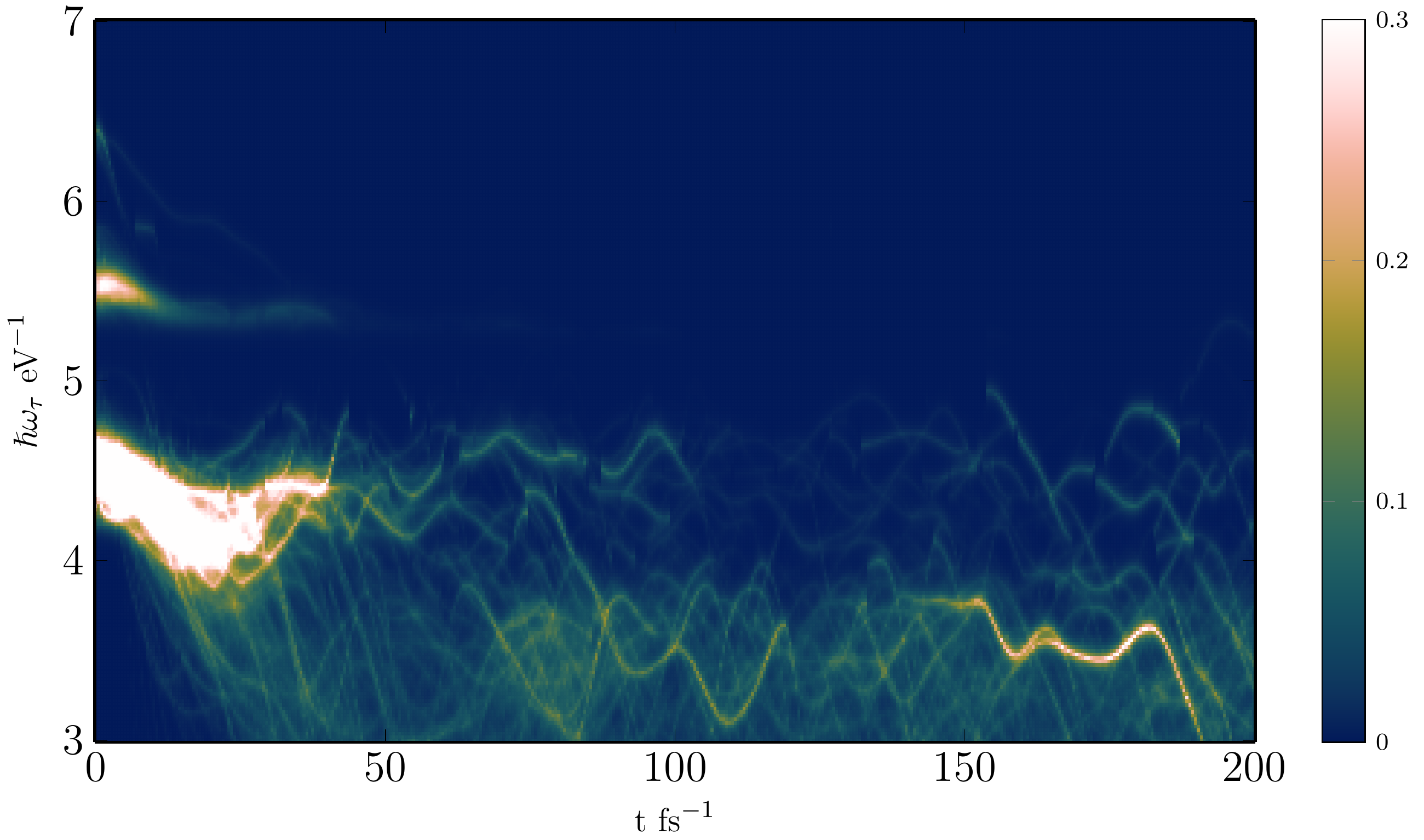}
			\put(10,52){\color{white}\captiontext*{}}
		\end{overpic}
    \end{subcaptionblock}
    \centering
    \begin{subcaptionblock}{.49\columnwidth}
        \centering
        \subcaptionlistentry{\(\kappa = 0.009\) and \(\gamma_{0} = 0.021 \)}
        \zlabel{fig:cav_es}
		\begin{overpic}[width=\textwidth]{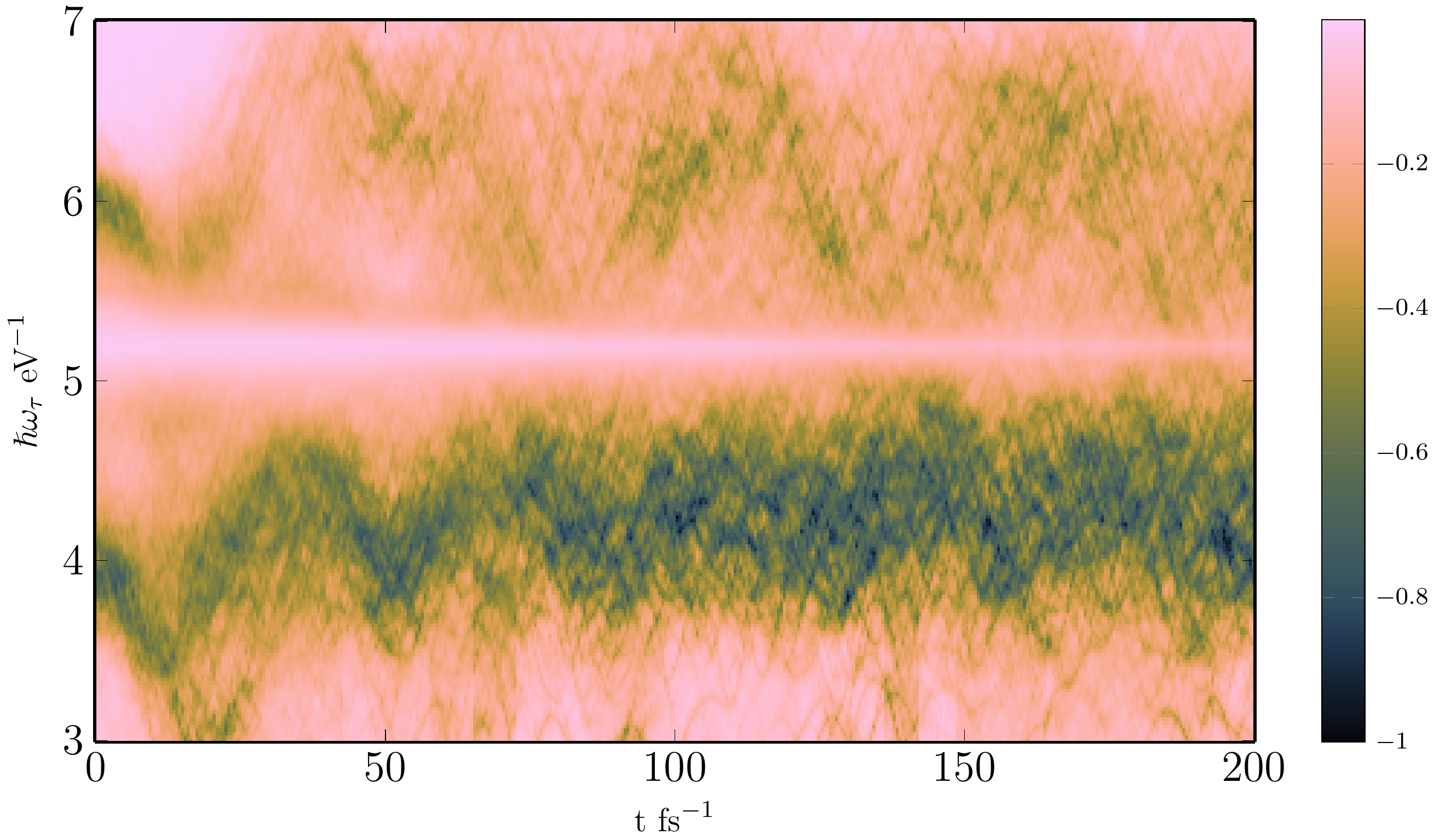}
			\put(10,52){\color{black}\captiontext*{}}
		\end{overpic}
    \end{subcaptionblock}
    \begin{subcaptionblock}{.49\columnwidth}
        \centering
        \subcaptionlistentry{\(\kappa = 0.025\) and \(\gamma_{0} = 0.011 \)}
        \zlabel{fig:cav_tt}
		\begin{overpic}[width=\textwidth]{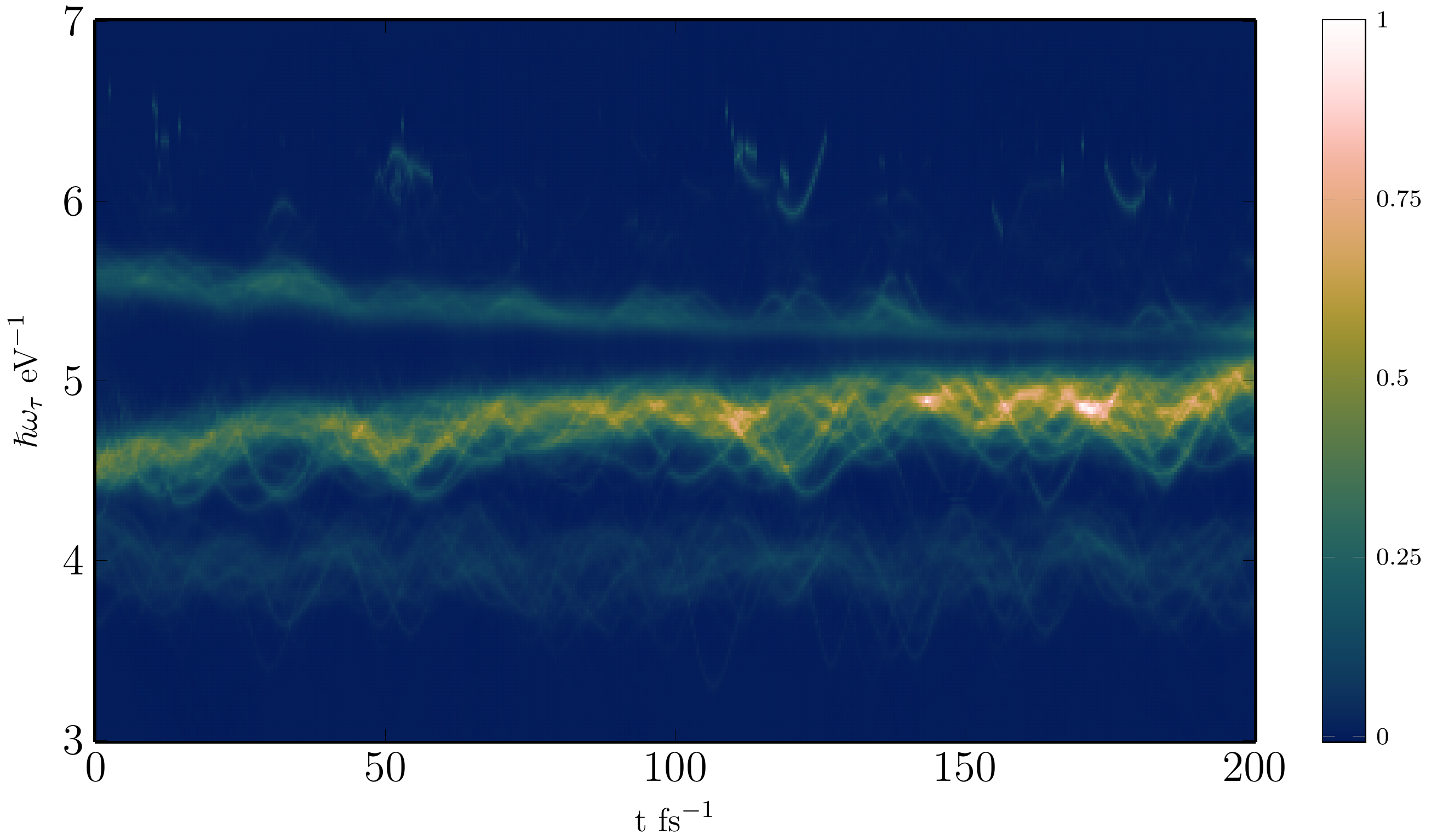}
			\put(10,52){\color{white}\captiontext*{}}
		\end{overpic}
    \end{subcaptionblock}
    \centering
    \caption{cQED TA PP signals (GSB, SE, and ESA) of pyrazine at resonance
        \emph{i.e.}\ \(\omega_{\textrm{c}} = \qty{5.2}{\eV}\).
        DW input values: pulse duration \(\tau = \qty{5.0}{\fs}\),
		pump frequency \(\omega_{\textrm{pu}} = \qty{5.2}{\eV}\), and 
		probe carrier frequencies \(\omega_{\textrm{pr}}\) in the
		range from \qtyrange{3.0}{7.0}{\eV} in steps of \qty{0.02}{\eV}.
		cQDE input values: cavity loss rate \(\kappa = \qty{0.018}{\per\fs}\) and the
		dipole coupling strength \(\gamma_{0} = \qty{0.56}{\eV}\).
    }
    \zlabel{fig:cqed_signals}
\end{figure}

The Rabi splitting stems from the nature of the
light-matter coupling as mapped from TA PP experiments. The GSB signal reflects
the primary eigenstates formed by the coupling, thus has the slightly larger
Rabi splitting. In contrast, the SE is slightly off-resonance with respect to
the GSB signal.
Therefore, one expects the excited states contributing to the SE to undergo
ultrafast radiationless decay on a timescale of approximately \qty{20}{\fs}, 
in the agreement with the pyrazine's TA PP spectra simulated without
cavity.\cite{Xiang21}
Finally, the ESA comprises transitions from the lower-lying states of manifold
\{I\} to higher-lying electronic states of manifold \{II\}. Because those
higher-lying electronic states
lie further from resonance with the cavity, those are less strongly coupled
to the cavity mode. This leads to a further narrowing of the Rabi splitting.

The cavity QED SE TA PP signal (\zcref{fig:cav_se}) exhibits characteristics
similar to its bare-molecular counterpart; however, the cavity QED SE ultrafast
non-radiative decay is approximately \qty{6}{\fs} faster, and lacks
bifurcation of the wavepackage in the
B\(_{\textrm{2u}} \left(\textrm{n}\pi^{\textrm{*}}\right) \) state. Additionally, the
revivals persist for longer durations. Beyond \qty{145}{\fs}, several spectral
signatures emerge connecting UP and LP, which intensify as the Rabi
splitting narrows.

We observe a spectral pattern characterized by an apparent overlap between
the intensities of the cavity QED GSB and ESA signals. Profile analysis of
the signals confirms this overlap (\zcref{sse:qed_profile} provides a
detailed comparison). Although this finding is expected
experimentally,\cite{McKillop2026} numerical simulations often struggle
to reproduce it, because this requires the explicit inclusion of coupling
across all simulated states. Nonetheless, the cQUEDA approach correctly
accounts for the decay from the ESA to the GSB. In the specific case of
the pyrazine chromophore, we observe GSB repopulation within \qty{22}{\fs}.

The total spectra (\zcref{fig:cqed_signals}) is mainly dominated by the GSB
signal. The second largest contribution arises from the ESA signal, as evidenced
by the small share of negative population in the colour bar (also seen
in \zcref{fig:profile_cavity}). A comparison of the total and GSB signals using
the mean squared absolute error reveals a difference of approximately \qty{5}{\percent}.

\section{Conclusions and perspective\zlabel{se:conclutions}}

In this work, we introduce a methodology for simulating cavity QED
time-resolved nonlinear signals within the framework of the quasi-classical
DW approximation. The methodology comprises two sets
of parameters: (a) cavity QED parameters that describe coupling of the molecular system
to the cavity, and (b) DW parameters that specify interaction of the system
wit the PP pulses. cQUEDA
takes quasi-classical surface-hopping trajectory simulations as input and
incorporates them to the cavity QED DW functions which generate TA PP signals.
We have implemented this methodology in the \texttt{WaveMixings.jl} package.

We demonstrate the applicability of the methodology by simulating the cavity QED
SE TA PP signal of pyrazine under four regimes: (a) no cavity coupling, (b) a
long-lived cavity, (c) moderate cavity coupling, and (d) a short-lived cavity.
In case (a), the signal reduces to the
standard integral TA PP signal in the DW approximation.\cite{Xiang21} In 
cases (b), (c), and (d), the signal reveals the presence of the UP and LP,
thereby confirming polariton formation. The SE TA PP signals in cases (c) and (d) are
particularly interesting. Under moderate cavity coupling in case (c), we observe
several over-excited trajectories, which come into resonance with the field
and therefore access the relevant polaritonic potential energy
surfaces of pyrazine. Although cases (b) and (d) appear similar at first glance
apart from the shorter cavity lifetime in (d), closer inspection of case (d) reveals
subtle differences. More importantly, after the UP and LP merge,
we observe increase in spectral intensity in several spectral regions.
These results confirm that the methodology does not bias the spectra and
faithfully reproduces nonlinear spectroscopic signals of molecular system.

We highlight two major contributions of this work to the field. First, to the
best of the authors knowledge, this is the first approach that simulates
nonlinear spectroscopic signals of cavity QED systems by \emph{ab initio}
methods. Although recent studies have shed light on the principles and
phenomena observed in linear spectroscopic signals, simulation of nonlinear
signals remain challenging.\cite{YuenZhou2024,Mukamel2011} Second, the
developed QED DW methodology contains a minimal set of the governing
parameters. Hence it can be conveniently extended in several directions.
For example, one can refine the input parameters describing the PP pulses
to better reproduce experimental signals.\cite{Deng2010,Thomas2024} 
Alternatively, one can refine the cavity parameters (for example, to
use multi-mode cavity field) to better capture the cavity QED physics. 
For example, one can implement a biexponential cavity-mode lifetime function
which would better reproduce the faster UP decay.

Recent state-of-the-art spectroscopic experiments have uncovered
fascinating cavity QED phenomena.\cite{Heimig2026,Kushida2025,Gutierrez2026,Hong2026}
The development of theoretical methods,
particularly master equation approaches, has made it possible to shed light on the
fundamental physics underlying these phenomena. Nevertheless, time-resolved nonlinear
spectroscopic signals have remained difficult to access and interpret. By introducing
the QED DW methodology, we aim to contribute to the exploration of this
largely challenging regime. With this goal in mind, here we outline a roadmap for
the ongoing and future development of the methodology. We are currently working to
relax the main assumption underlying in our methodology, which is achievable
by performing quasi-classical trajectory simulations of coupled light-matter
dynamics. Another promising line of development is the extension of the
formalism to electronic two-dimensional (2D) 
spectroscopy. Although master equation approaches already developed methodologies
for the simulation of 2D cavity QED spectroscopy,\cite{Mondal2025} it is promising 
to develop an on-the-fly QED DW framework for the evaluation of polaritonic 2D spectra.

Finally, because of the intrinsic nature of cavity loss and its mathematical
formulation in light-matter interactions, many quasi-classical simulations
either neglect it or account for it only through \emph{a posteriori} corrections.
In fact, numerical studies of cavity loss remain somewhat inconclusive: some
approaches suggest that it has a significant effect, whereas others indicate
that its influence is practically negligible. In light of this,
it would be valuable to incorporate suitable functions into cQUEDA to account
for cavity loss \emph{a posteriori} and consequently
clarify its role more systematically.

\section*{Acknowledgements}

We thank Professor Wolfgang Domcke for its helpful commentaries that
improved the manuscript. L.V. acknowledges support from the Hosei University
International Fellowship. K. S. acknowledges support from the National Natural Science
Foundation of China (Grant No. 22573025). M. F. G. acknowledges support
from the National Natural Science Foundation of China (No. 22373028).

\section*{Supporting information}

Find implementation of the cQUEDA methodology in the \texttt{WaveMixings.jl}
repository (\url{https://codeberg.org/apolionl/WaveMixings.jl}).

\appendix

\section{Derivation of the cQUEDA\zlabel{sec:qed_derive}}

\begin{IEEEeqnarray}{rCl}
    D(\omega_{p}) & = &
    \int_{-\infty}^{\infty}dt_{2}'\int_{0}^{\infty}dt_{1}E_{p}(t_{2}')E_{p}(t_{2}'-t_{1})
    e^{i\omega_{p}t_{1}}e_{\rightarrow}^{-i\int_{0}^{t_{1}}\mathcal{H}_{\mathrm{I}}(t')dt'}
    \rho_{0}e^{iH_{0}t_{1}}+H.c.
    \IEEEeqnarraynumspace
    \zlabel{eq:dens}
\end{IEEEeqnarray}
If
\begin{IEEEeqnarray}{rCl}
    \Delta(t) & = &
    \begin{cases}
        \Delta,  & t\leq\kappa \\
        0,       & t>\kappa
    \end{cases}
    \IEEEeqnarraynumspace
    \zlabel{eq:limits}
\end{IEEEeqnarray}
then is centred around $t=0$, then
\begin{IEEEeqnarray}{rCl}
    e_{\rightarrow}^{-i\int_{0}^{t_{1}}\mathcal{H}_{\mathrm{I}}(t')dt'} & = &
    \begin{cases}
        e^{-i\mathcal{H}_{\mathrm{I}}t_{1}},                                                          & t_{1}\leq\kappa \\
        e^{-i\mathcal{H}_{\mathrm{I}}\kappa}e^{-i\overline{\mathcal{H}}_{\mathrm{I}}(t_{1}-\kappa)},  & t_{1}>\kappa
    \end{cases}
    \IEEEeqnarraynumspace
    \zlabel{eq:limits_rot}
\end{IEEEeqnarray}
where
\begin{IEEEeqnarray}{rCl}
    \mathcal{H}_{\mathrm{I}} =
    \begin{pmatrix}
        H_{\mathrm{I}} & \Delta         \\
        \Delta         & H_{\mathrm{c}}
    \end{pmatrix},
    & \ \ \  &
    \overline{\mathcal{H}}_{\mathrm{I}} =
    \begin{pmatrix}
        H_{\mathrm{I}} & 0              \\
        0              & H_{\mathrm{c}}
    \end{pmatrix}. \nonumber
    \IEEEeqnarraynumspace
    \zlabel{eq:minim_ham_dw}
\end{IEEEeqnarray}
Then
\begin{IEEEeqnarray}{rCl}
    e^{-i\mathcal{H}_{\mathrm{I}}\kappa} & = &
    \sum_{\alpha=1,2} \ket{\alpha}\bra{\alpha} e^{-iE_{\alpha}\kappa}
    \IEEEeqnarraynumspace
    \zlabel{eq:rot_0}
\end{IEEEeqnarray}
\begin{IEEEeqnarray}{rCl}
    e^{-i\overline{\mathcal{H}}_{\mathrm{I}}(t_{1}-\kappa)} & = &
    \sum_{\beta=e,c} \ket{\beta}\bra{\beta} e^{-iE_{\alpha}(t_{1}-\kappa)}.
    \IEEEeqnarraynumspace
    \zlabel{eq:rot_I}
\end{IEEEeqnarray}

Since the cavity mode is initially in the state $\ket{c}$ (the
cavity mode is excited) we need to calculate
\begin{IEEEeqnarray}{rCl}
    \bra{c}e^{-i\mathcal{H}_{\mathrm{I}}\kappa}e^{-i\overline{\mathcal{H}}_{\mathrm{I}}(t_{1}-\kappa)}\ket{c} & = &
    \bra{c}\sum_{\alpha=1,2}\ket{\alpha}\bra{\alpha} e^{-iE_{\alpha}\kappa}\sum_{\beta=e,c}\ket{\beta}\bra{\beta} e^{-iE_{\beta}(t_{1}-\kappa)}\ket{c}
    \IEEEeqnarraynumspace
    \zlabel{eq:cav_c}
\end{IEEEeqnarray}
\begin{IEEEeqnarray}{rCl}
    \bra{c}e^{-i\mathcal{H}_{\mathrm{I}}\kappa}e^{-i\overline{\mathcal{H}}_{\mathrm{I}}(t_{1}-\kappa)}\ket{c} & = &
    \bra{c}\sum_{\alpha=1,2}\ket{\alpha}\bra{\alpha} e^{-iE_{\alpha}\kappa}
    \sum_{\beta=e,c}\ket{\beta}\bra{\beta} e^{-iE_{\beta}(t_{1}-\kappa)}\ket{c} \nonumber \\
    & = &
    \bra{c}\sum_{\alpha=1,2}\ket{\alpha}\bra{\alpha} e^{-iE_{\alpha}\kappa}\ket{c} e^{-iE_{c}(t_{1}-\kappa)}  \nonumber \\
    & = &
    \left(\left|\braket{c|1}\right|^{2}e^{-iE_{1}\kappa}+\left|\braket{c|2}\right|^{2}e^{-iE_{2}\kappa}\right)e^{-iE_{c}(t_{1}-\kappa)}
    \nonumber \\
    & = & \left(\left|\braket{c|1}\right|^{2}e^{-i(E_{1}-E_{c})\kappa} +\left|\braket{c|2}\right|^{2}e^{-i(E_{2}-E_{c})\kappa}\right)e^{-iE_{c}t_{1}}
    \IEEEeqnarraynumspace
    \zlabel{eq:cav_cc}
\end{IEEEeqnarray}

We thus have
\begin{IEEEeqnarray}{rCl}
    \bra{c} e_{\rightarrow}^{-i\int_{0}^{t_{1}}\mathcal{H}_{\mathrm{I}}(t')dt'}	\ket{c}
    & = &
    \begin{cases}
        \left( \left|\braket{c|1}\right|^{2}e^{-iE_{1}t_{1}}+ \left|\braket{c|2}\right|^{2}e^{-iE_{2}t_{1}}\right),                                      & t_{1}\leq\kappa \\
        \left( \left|\braket{c|1}\right|^{2}e^{-i(E_{1}-E_{c})\kappa}+ \left|\braket{c|2}\right|^{2}ee^{-i(E_{2}-E_{c})\kappa}\right) e^{-iE_{c}t_{1}},  & t_{1}>\kappa
    \end{cases}
    \IEEEeqnarraynumspace
    \zlabel{eq:brac_ket_lims}
\end{IEEEeqnarray}

\begin{IEEEeqnarray}{rCl}
    \ket{1} & = & \cos(\phi)\ket{c}+\sin(\phi)\ket{e}
    \IEEEeqnarraynumspace
    \zlabel{eq:cos}
\end{IEEEeqnarray}
\begin{IEEEeqnarray}{rCl}
    \ket{2} & = & \sin(\phi)\ket{c}+\cos(\phi)\ket{e}
    \IEEEeqnarraynumspace
    \zlabel{eq:sin}
\end{IEEEeqnarray}
It seems that the idea of averaging over $\ket{c}$ is incorrect.
We start from the global vacuum state, then the pump pulse excites
the system and when it gets excited it splits due to the coupling
with the vacuum field. Then I think we need to calculate
\begin{IEEEeqnarray}{rCl}
    \bra{e} e_{\rightarrow}^{-i\int_{0}^{t_{1}}\mathcal{H}_{\mathrm{I}}(t')dt'}	\ket{e}
    & = &
    \begin{cases}
        \left( \left|\braket{e|1}\right|^{2}e^{-iE_{1}t_{1}}+ \left|\braket{e|2}\right|^{2}e^{-iE_{2}t_{1}}\right),                                     &   t_{1}\leq\kappa \\
        \left( \left|\braket{e|1}\right|^{2}e^{-i(E_{1}-E_{c})\kappa}+ \left|\braket{e|2}\right|^{2}ee^{-i(E_{2}-E_{c})\kappa}\right) e^{-iE_{c}t_{1}}, &   t_{1}>\kappa
    \end{cases}
    \nonumber \\
    & = &
    \begin{cases}
        \left(\sin^{2}(\phi)e^{-iE_{1}t_{1}}+\cos^{2}(\phi)e^{-iE_{2}t_{1}}\right),                                    & t_{1}\leq\kappa \\
        \left(\sin^{2}(\phi)e^{-i(E_{1}-E_{e})\kappa}+\cos^{2}(\phi)e^{-i(E_{2}-E_{e})\kappa}\right)e^{-iE_{e}t_{1}},  & t_{1}>\kappa
    \end{cases}
    \IEEEeqnarraynumspace
    \zlabel{eq:averag_e}
\end{IEEEeqnarray}

Thus, if $\kappa\ll1$, we obtain
\begin{IEEEeqnarray}{rCl}
    \bra{e} e_{\rightarrow}^{-i\int_{0}^{t_{1}}\mathcal{H}_{\mathrm{I}}(t')dt'}\ket{e} & = & e^{-iE_{e}t_{1}}
    \IEEEeqnarraynumspace
    \zlabel{eq:tau_c_small}
\end{IEEEeqnarray}
that is nothing happens. If $\,t_{1}\sim\kappa$, we can neglect
the cavity coupling between the pulses and write 
\begin{IEEEeqnarray}{rCl}
    \bra{e} e_{\rightarrow}^{-i\int_{0}^{t_{1}}\mathcal{H}_{\mathrm{I}}(t')dt'}\ket{e} & = & e^{-iE_{1}t_{1}}\cos^{2}(\phi)+e^{-iE_{2}t_{1}}\sin^{2}(\phi)
    \IEEEeqnarraynumspace
    \zlabel{eq:tau_c_small_corr}
\end{IEEEeqnarray}
which is a clear signature of the left and right excitons. Interestingly,
it is independent of $\kappa$.

The setting is that the photonic mode switches on at time $t=0$.

Let
\begin{IEEEeqnarray}{rCl}
    \Omega  =  \sqrt{(E_{c}-E_{e})^{2}+4\xi^{2}}, & \ \ \ & \sigma=\Omega+E_{c}-E_{e}. \nonumber
    \IEEEeqnarraynumspace
    \zlabel{eq:rabi_der}
\end{IEEEeqnarray}
Then
\begin{IEEEeqnarray}{rCl}
    \cos^{2}(\phi) = \frac{2\xi^{2}}{\Omega\sigma}, & \ \ \ & \sin^{2}(\phi)=\frac{\sigma}{2\Omega} \nonumber
    \IEEEeqnarraynumspace
    \zlabel{eq:trig_der}
\end{IEEEeqnarray}
\begin{IEEEeqnarray}{rCl}
    A_{eg}(\omega_{pu},0,\boldsymbol{R}_{g}) &=& \sum_{k=1,2}E_{pu}^{2}(\omega_{pu}-U_{eg}^{(k)}(0,\boldsymbol{R}_{g}))\eta_{k}(0,\boldsymbol{R}_{g})
    \IEEEeqnarraynumspace
    \zlabel{eq:dw0_der}
\end{IEEEeqnarray}
\begin{IEEEeqnarray}{rCl}
    B_{e(T)f}(\omega_{pu},T,\boldsymbol{R}_{e}(T)) & = &\sum_{k=1,2}E_{pu}^{2}(\omega_{pu}-U_{e(T)g}^{(k)}(T,\boldsymbol{R}_{e}(T)))\eta_{k}(T,\boldsymbol{R}_{e}(T))
    \IEEEeqnarraynumspace
    \zlabel{eq:dwI_der}
\end{IEEEeqnarray}

\clearpage

\end{document}